\documentclass{aa}
\usepackage{graphicx}
\usepackage{txfonts}
\usepackage{hyperref}
\usepackage[normalem]{ulem}
\hypersetup{colorlinks = true, citecolor = {blue}, urlcolor= {blue}}

\def\gapprox{\;\rlap{\lower 3.0pt                       
        \hbox{$\sim$}}\raise 2.5pt\hbox{$>$}\;}
\def\lapprox{\;\rlap{\lower 3.1pt                       
        \hbox{$\sim$}}\raise 2.7pt\hbox{$<$}\;}

\newcommand{\be}{ \begin{equation} }
\newcommand{\ee}{\end{equation}}

\newcommand{\ben}{\begin{enumerate}}
\newcommand{\een}{\end{enumerate}}

\renewenvironment{thebibliography}[1]{\begin{oldthebibliography}{#1}\setlength{\parskip}{0ex}\setlength{\itemsep}{0ex}}{\end{oldthebibliography}}

\usepackage{diagbox}
\usepackage{siunitx}
\newcommand{\orcid}[1]{\href{https://orcid.org/#1}{\protect\includegraphics[width=8pt]{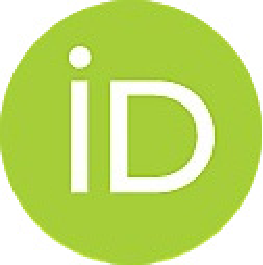}}}
\makeatletter
\renewcommand*\aa@pageof{, page \thepage{} of \pageref*{LastPage}}
\makeatother

\usepackage[dvipsnames]{xcolor}

\begin{document}
   
\title{Milky Way globular clusters on cosmological timescales. IV. Guests in the outer Solar System}

\author{Maryna~Ishchenko
\inst{1, 2, 3}\orcid{0000-0002-6961-8170 }
\and
Peter~Berczik
\inst{1, 2, 3, 4}\orcid{0000-0003-4176-152X}
\and
Margarita~Sobolenko
\inst{1, 2}\orcid{0000-0003-0553-7301}
}

\institute{Main Astronomical Observatory, National Academy of Sciences of Ukraine,
27 Akademika Zabolotnoho St, 03143 Kyiv, Ukraine           
\email{\href{mailto:marina@mao.kiev.ua}{marina@mao.kiev.ua}}
\and
Nicolaus Copernicus Astronomical Centre Polish Academy of Sciences, ul. Bartycka 18, 00-716 Warsaw, Poland
\and
Fesenkov Astrophysical Institute, Observatory 23, 050020 Almaty, Kazakhstan
\and
Konkoly Observatory, Research Centre for Astronomy and Earth Sciences, E\"otv\"os Lor\'and Research Network (ELKH), MTA Centre of Excellence, Konkoly Thege Mikl\'os \'ut 15-17, 1121 Budapest, Hungary
}
   
\date{Received xxx / Accepted xxx}
    
\abstract    
{The present epoch of the \textit{Gaia} success gives us a possibility to predict the dynamical evolution of our Solar System in the global Galactic framework with high precision.}
{We statistically investigated the total interaction of globular clusters with the Solar System during six billion years of look-back time. We estimated the gravitational influence of globular clusters' flyby onto the Oort cloud system.}
{To perform the realistic orbital dynamical evolution for each individual cluster, we used our own high-order parallel dynamical $N$-body $\varphi$-GPU code that we developed. To reconstruct the orbital trajectories of clusters, we used five external dynamical time variable galactic potentials selected from the IllustrisTNG-100 cosmological database and one static potential. To detect a cluster's close passages near the Solar System, we adopted a simple distance criteria of below 200~pc. To take into account a cluster's measurement errors (based on \textit{Gaia} DR3), we generated 1000 initial positions and velocity randomisations for each cluster in each potential.}
{We found 35 globular clusters that have had close passages near the Sun in all the six potentials during the whole lifetime of the Solar System. We can conclude that at a relative distance of 50~pc between a GC and the SolS, we obtain on average $\sim 15$\% of the close passage probability over all six billion years, and at $dR=100$~pc, we get on average $\sim 35$\% of the close passage probability over all six billion years. The globular clusters BH~140, UKS~1, and Djorg~1 have a mean minimum relative distance to the Sun of 9, 19, and 17~pc, respectively. We analysed the gravitational energetic influence on the whole Oort cloud system from the closest selected globular cluster flyby. We generally found that a globular cluster with a typical mass above a few times $10^{5} {rm M}_\odot$ and with deep close passages in a 1--2 pc immediately results in the ejection more than $\sim$ 30\% of particles from the Oort cloud system.}
{We can assume that a globular cluster with close passages near the Sun is not a frequent occurrence but also not an exceptional event in the Solar System's lifetime.}


\titlerunning{IV. Guests in the outer Solar System}
\authorrunning{M.~Ishchenko et al.}
\maketitle

\section{Introduction}\label{sec:Intr}

Our Solar System (SolS) is located in a relatively calm region of the Galaxy, namely, between two spiral arms (Perseus and Scutum-Centaurus) at a distance of $\approx8$~kpc from the centre of the Galaxy. During the analysis of globular cluster (GC) orbits in our previous paper \cite{Ishchenko2023a} (hereafter, \hyperlink{I23}{\color{blue}{Paper~I}}), we found that more than 50~GCs can potentially cross the orbital trajectory of the Sun. In our following paper, we concluded that even a 'collision' between two GCs is not a unique phenomenon  \citep{Ishchenko2023c}. Globular clusters are objects that can even actively interact with the Galactic centre \citep{Ishchenko2023b}, open clusters \citep{Marcos2014}, and with each other \citep{Khoperskov2018, Ishchenko2023c}. Thus, we assume that close passages of individual stars could have a gravitational influence not only on SolS dynamics \citep{Marcos2022, Portegies2021a, Portegies2021b, Bailer-Jones2022} and star clusters \citep{Pfalzner2020, Pfalzner2021, Pfalzner2021a} but potentially also on a whole GC as well.

In this paper, our aim is to investigate the possible gravitational influence of close passages of GCs on the SolS. Our main tasks can be formulated as follows: performing statistical analysis of the total interaction of GCs with the SolS, estimating the interaction and probability of close passages near the SolS for individual GCs, and estimating the gravitational influence onto the Oort cloud due to the potential close passages of GCs.

To add more realism to our dynamical orbital evolution (simulations), we added to the standard $N$-body code the time-dependent external potentials, which were selected from IllustrisTNG-100 cosmological simulation database \citep{Nelson2018, Nelson2019, NelsonPill2019}. We selected five IllustrisTNG time variable potentials (TNG-TVP) that (at present redshift $z = 0$) have parameters (halo and disc masses, and their characteristic scales) similar to our present day Galaxy: {\tt 411321, 441327, 451323, 462077,} and {\tt 474170}. The procedure for sampling and fitting the selected potentials is described in detail in the papers \hyperlink{I23}{\color{blue}{Paper~I}} and \cite{Mardini2020}. The code routines are also publicly available at GitHub.\footnote{The ORIENT: \\~\url{ https://github.com/Mohammad-Mardini/The-ORIENT}} As a typical example, we present the main parameters (halo and disc masses, and their characteristic scales) for {\tt 411321} TNG-TVP in Table~\ref{tab:pot}. For a more detailed description of all five selected TNG-TVP, we refer to our previous \hyperlink{I23}{\color{blue}{Paper~I}}.

\begin{table}[htbp!]
\caption{Parameters of the potentials at redshift zero. The last column shows the parameters of the corresponding Milky Way components according to \cite{Bennett2022} and \cite{Bland-Hawthorn2016}.}
\centering
\resizebox{0.49\textwidth}{!}{
\begin{tabular}{llccc}
\hline
\hline
\multicolumn{1}{c}{Parameter} & Unit & {\tt 411321} & {\tt FIX} & Milky Way \\
\hline
\hline
Bulge mass, $M_{\rm d}$         & $10^{10}~\rm M_{\odot}$ & -- & 1.4 & $\sim1.4$ \\
Disc mass,  $M_{\rm d}$         & $10^{10}~\rm M_{\odot}$ & 7.110 & 9.0 & 6.788 \\
Halo mass,  $M_{\rm h}$         & $10^{12}~\rm M_{\odot}$ & 1.190 & 0.72 & 1.000 \\

Bulge scale length, $a_{\rm d}$ & 1~kpc                     & -- & 0.0 & $\sim0.0$ \\
Bulge scale height, $b_{\rm d}$ & 1~kpc                     & -- & 0.3 & $\sim1.0$ \\

Disc scale length, $a_{\rm d}$ & 1~kpc                     & 2.073 & 3.3 & 3.410 \\
Disc scale height, $b_{\rm d}$ & 1~kpc                     & 1.126 & 0.3 & 0.320 \\
Halo scale height, $b_{\rm h}$ & 10~kpc                    & 2.848 & 2.5 & 2.770 \\
\hline
\end{tabular}
}
\label{tab:pot}
\end{table} 

In addition to our five selected TNG-TVPs, we added another fixed potential whose parameters (such as halo, disc, and bulge masses, and their characteristic scales) do not change over time. We refer to this potential using the code-name 'FIX'. For this potential in our simulations, we used the three-component (bulge-disc-halo) axisymmetric Plummer-Kuzmin model \citep{Miyamoto1975}, as in  \cite{Just2009} and \cite{Shukirgaliyev2021}. We present the parameters (halo and disc masses, and their characteristic scales) for TNG external and static potentials in Table~\ref{tab:pot}.

As an initial condition for our GC system, we used the randomised 6D phase-space orbital information provided from \cite{Baumgardt2021}.\footnote{\label{note1}Error values for D$_{\odot}$, pmRA, pmDEC, and RV from \url{https://people.smp.uq.edu.au/HolgerBaumgardt/globular/orbits_table.txt}} Using 1000 initial data sets for each potential, in total we obtained 6000 runs for the GC systems in the integration procedure. More details about the dynamical integration can be found in our previous papers, \cite{Ishchenko2023b} and \cite{Ishchenko2023c}). 

To reproduce the orbit of the Sun in the past, we accepted an in-plane distance of the Sun from the Galactic centre at the plane as $X_{\odot} = 8.178$~kpc \cite{Gravity2019}, $Z_{\odot} = 20.8$~pc \cite{Bennett2019}. The velocity transformation is basically described in \cite{Johnson1987}, but for the equatorial position of the North Galactic Pole (NGP), we used the updated values from \cite{Karim2017}: RA$_{\rm NGP}  = 192\fdg7278$, DEC$_{\rm NGP} =  26\fdg8630$, $\theta_{\rm NGP}  = 122\fdg9280$. For the Local Standard of Rest~(LSR) velocity, we set $V_{\rm LSR} = 234.737$~km~s$^{-1}$ \citep{Reid2004}, and for the peculiar velocity of the Sun with respect to the LSR, we used $U_{\odot} = 11.1$~km~s$^{-1}$, $V_{\odot} = 12.24$~km~s$^{-1}$, $W_{\odot} = 7.25$~km~s$^{-1}$~\citep{Schonrich2010}. Since the integration was performed backwards in time, the sign of the velocity components for Sun and GCs were changed to the opposite value. 

Using these observations as initial conditions, we performed simulations of 147~GCs together with the Sun up to six billion years in look-back time. As we know, the age of the Sun is $\sim5$~Gyr since it is set on the main sequence \citep{Bonanno2002, Connelly2012S}. We added an additional 1~Gyr in order to take into account some dynamical time for star formation itself.  

The paper is organised as follows. In Section~\ref{sec:rate}, we present the total interaction of the GCs with the SolS. In Section~\ref{sec:ind_gc}, we present the individual analysis of GC interaction. In Sections~\ref{sec:force} and \ref{sec:con}, we present the gravity influence of GCs to the Oort cloud system and summarise our findings.

\section{Total interaction of globular clusters with the Solar System}\label{sec:rate}

As a first step, we estimated all the close passage interactions of the GCs with the SolS. We used only one main criterion to characterise the close passages $N_{\rm pass}$ between the GCs and SolS, namely, the separation $dR$ between the GC and SolS should be less than 200~pc. The close passages of typical GCs with masses on order of $\sim10^5\rm\;M_\odot$ at such a distance can already have a strong gravitational influence on the outer part of the SolS \citep{Portegies2021a, Portegies2021b}.

By analysing the results from all 6000~runs with the close passages that correspond to our criteria, we found only 56~GCs that have close passages near the SolS in at least one of the random realisations in one of the six selected potentials. Of these GCs, only 35 have close passages near the SolS in all the six potentials. Under our assumptions and taking into account the fact that in the used \textit{Gaia} DR3 sample we have 147~GCs, in total we found that 38\% of the 147 GCs could potentially have had close passages near the SolS during the whole six-billion-year evolution.

\begin{figure*}[htbp!]
\centering
\includegraphics[width=0.49\linewidth]{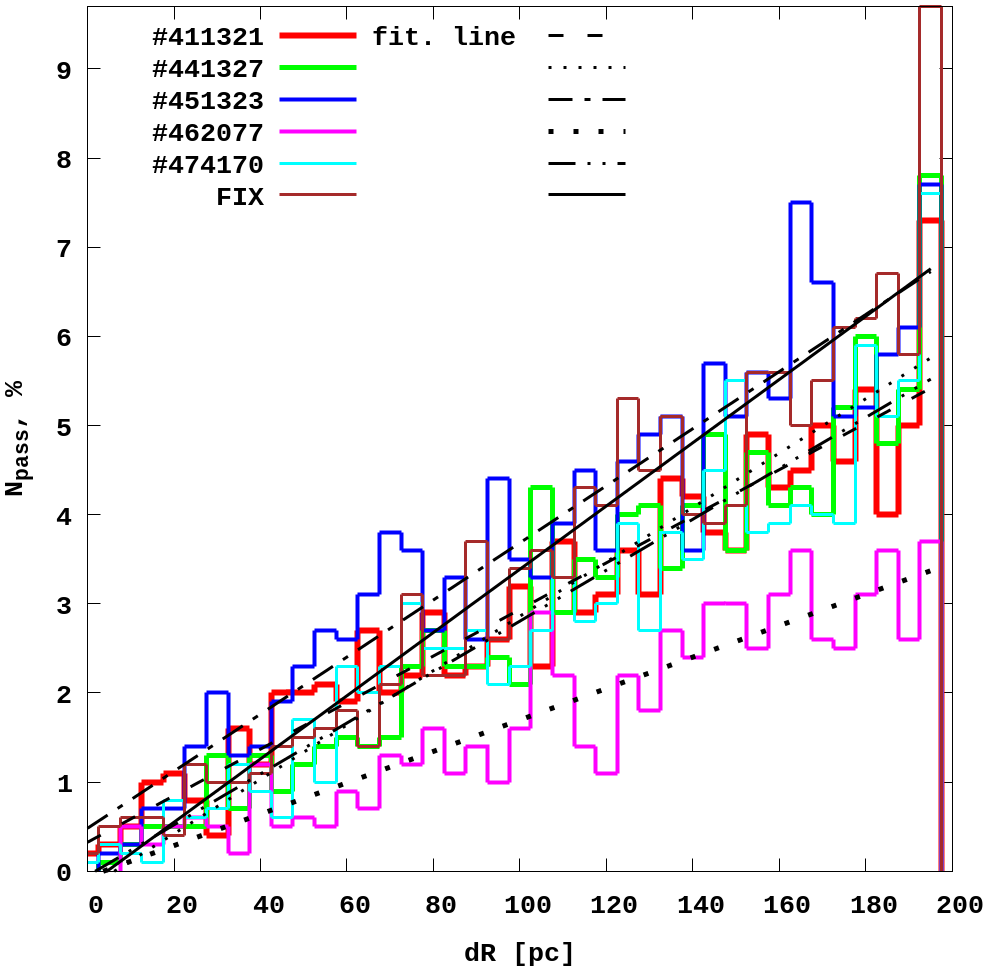}
\includegraphics[width=0.49\linewidth]{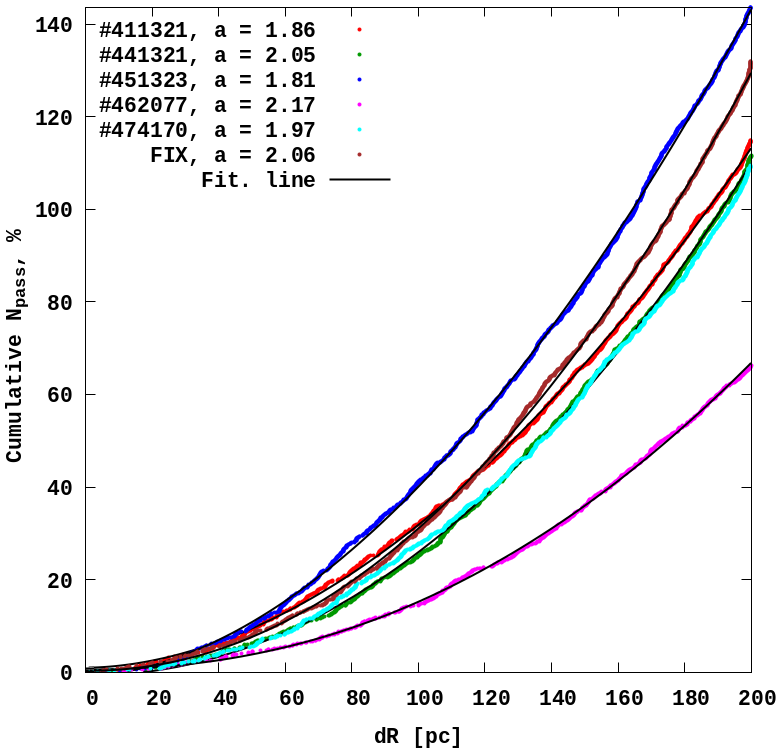}
\caption{Average probability of the number of passages (left panel) between GCs and SolS as a function of the relative distance $dR$. The black dashed lines are the linear fit for the corresponding potentials. The right panel presents the average probability of the cumulative number of passages as a function of the minimum relative distance $dR$  of the GCs at the moment of closest approach in all six potentials. Colours: {\tt 411321} -- red, {\tt 441327} -- green, {\tt 451323} -- blue, {\tt 462077} -- magenta, and {\tt 474170} -- cyan. The brown colour is for the FIX potentials.}
\label{fig:dr_fit}
\end{figure*}

In Fig.~\ref{fig:dr_fit} (left panel), we present the close passage interaction between GCs and SolS $N_{\rm pass}$ as a function of the relative distance $dR$ for all the six potentials with all the individual randomisations for all 6000~runs in total. We present each potential in a different colour and averaged them over all the 1000~randomisations). The black dashed lines represent the linear fit for the corresponding potentials. Thus, we observed that the close passage interaction increases linearly with the relative distance between the GCs and SolS. The largest close passage interaction ($N_{\rm pass}$) in the six billion years was obtained for {\tt 451323} TNG-TVP and in FIX potentials (blue and brown colours). We found the minimum number of close passages with the {\tt 462077} TNG-TVP potential (magenta colour). 

In Fig.~\ref{fig:dr_fit} (right panel), we present the average probability of the cumulative number of passages as a function of the GCs' minimum relative distance $dR$ at the moment of closest approach to the SolS in all the six potentials (summed up for the 1000~randomisations). As can be seen, the resulting distribution as a function of $dR$ is well fit by the simple power-law function:
\begin{equation}
N_{\rm pass}(dR)={\rm b} \cdot (dR)^{\rm a}, 
\label{eq:fit}
\end{equation}
where $\rm a = 1.99\pm0.13$ is the average best-fit slope parameter among our six variants of external potentials.

\begin{figure*}[htbp!]
\centering
\includegraphics[width=0.49\linewidth]{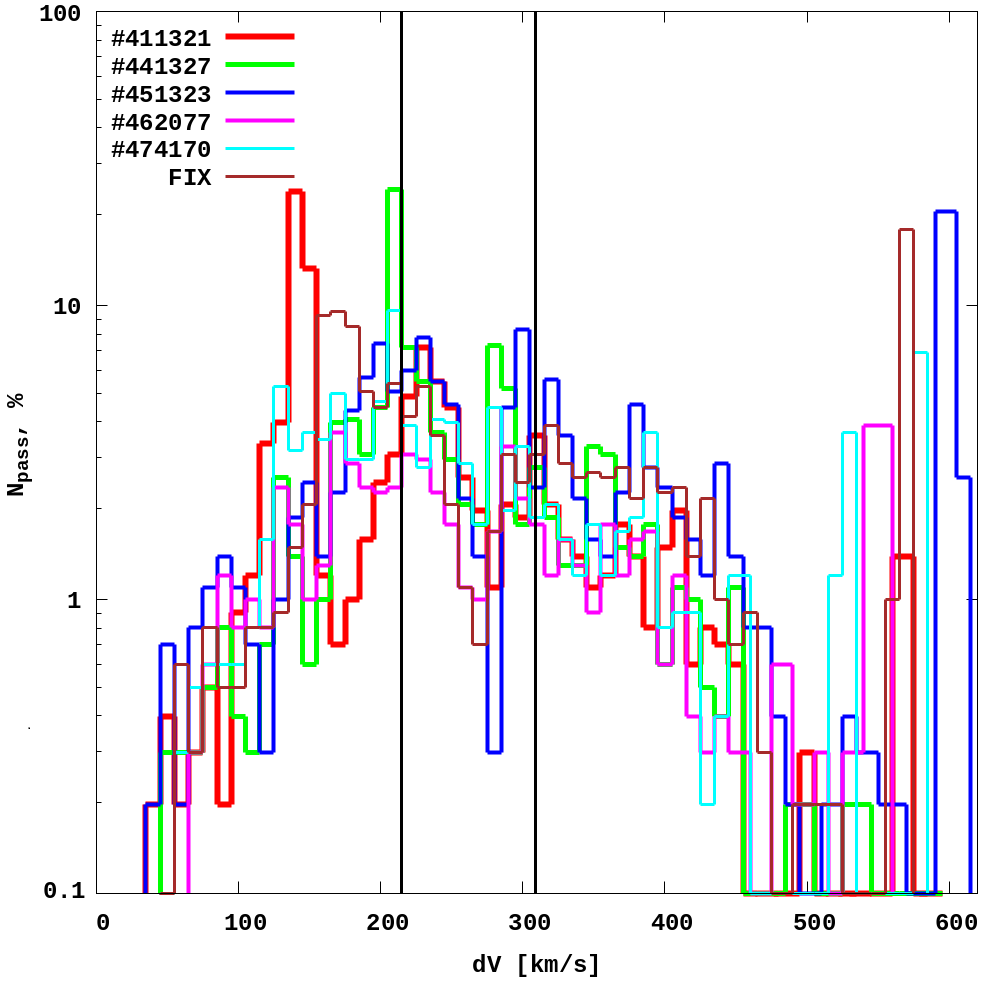}
\includegraphics[width=0.49\linewidth]{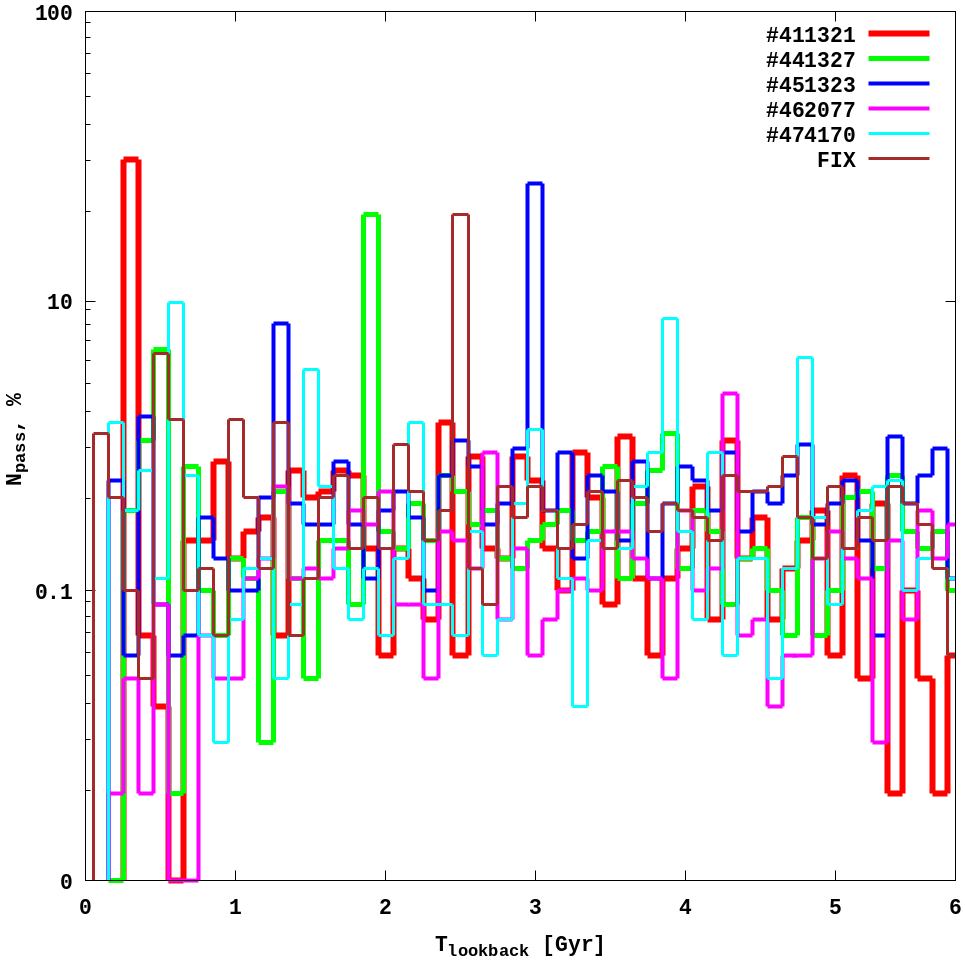}
\caption{Relative velocities (left panel) and time when close passages happen (right panel) between the GCs and the SolS. Colours: {\tt 411321} -- red, {\tt 441327} -- green, {\tt 451323} -- blue, {\tt 462077} -- magenta, and {\tt 474170} -- cyan. The brown colour is for the FIX potentials.}
\label{fig:dv_dt}
\end{figure*}

As can be seen, the of the cumulative close passage can be described by a simple quadratic relation: $N_{\rm pass} \sim (dR)^{2}$. By analysing Fig.~\ref{fig:dr_fit} (right panel), for example, we can conclude that at a relative distance of 50~pc between a GC and the SolS, we get on average $\sim 15$\% of the close passage probability over all six billion years, and at $dR=100$~pc, we get on average $\sim 35$\% of the close passage probability over all six billion years. In addition, Fig.~\ref{fig:dv_dt} (right panel) presents the distribution of GC close passages $N_{\rm pass}$ near the SolS for all six potentials as a function of look-back time. The average probability, around 2--3\%, is almost stable throughout the entire period of six billion years, except for the increase at the beginning (0--0.7~Gyr) and the decrease at the end (5.5--6~Gyr) of the integration. 

The main differences between the close encounter passages in Fig.~\ref{fig:dr_fit} and the different external potentials can be explained based on the slightly different behaviour of our TNG-TVP potentials' mass and size time evolution. For example, the encounter results for the potential {\tt 462077} are significantly lower compared to the other potentials. This can be explained as being due to the specific behaviour of the disc component scale lengths (a$_{\rm d}$ and b$_{\rm d}$) in our last six billion years of look-back time (\hyperlink{I23}{\color{blue}{Paper~I}}, see Fig.~2 therein). Interested readers can find a more detailed visualisation of all 159 GC orbits for different TNG-TVP potentials on the website of the project. \footnote{\label{note2}\url{https://bit.ly/3b0lafw}}

In Fig.~\ref{fig:dv_dt} (left panel), we present the velocity distribution for GCs that have close passages near the SolS. As can be seen, we found two groups with the first maximum around $\sim200$~km~s$^{-1}$ and the second near the $\sim300$~km~s$^{-1}$ (marked with black lines), where we have $\sim7$\% of the probability of close passages. The main contribution in the first peak comes mainly from three clusters, BH~140, Djorg~1, and UKS~1, which have a close passage probability of more than $\sim$ 20\%, on average, for the potentials. In the second peak, we have mainly the influence from the GCs that are in italics in the first block in Table~\ref{tab:GC_SS}.  

\section{Individual analysis of the interaction of globular clusters with the Solar System}\label{sec:ind_gc}

As a second step, we estimated the individual GC interaction with the SolS. In Table~\ref{tab:GC_SS}, we present GCs that have close passages near the SolS divided across three blocks. In the first block, we present the selected 35~GCs that have close passages in all six potentials. In the second block are 8~GCs that have close passages in five potentials, and in the third are 6~GCs that have close passages only in four potentials. Also, in Table~\ref{tab:GC_SS}, we present the kinematic parameters and properties such as the average minimum distance between the GC and the SolS ($<dR_{\rm m}>$) in all potentials and their relative mean velocities at the moment of collision ($<dV>$). In Table~\ref{tab:GC_SS}, we also present the current masses of the GCs and the half-mass radii as well as the type of orbit (TO), association of the Galaxy region (GR), and possible progenitor. The GCs IC~1276, NGC~5279, NGC~5466, NGC~6101, NGC~6333, NGC~4590, NGC~5034, and Pal~14 (eight~GCs in total) are not presented in the Table~\ref{tab:GC_SS} due to their extremely low collision probability (less than 0.1\%) with the SolS and the full absence of such collisions in three or more potentials. In next paragraph, we present several conclusions about the selected GCs that have close passages near the SolS.
    
The GCs BH~140, UKS~1, and Djorg~1 have a mean minimum relative distance value of $dR_{\rm m}$: $9\pm5$, $19\pm6$ and $17\pm9$~pc between the GC and the SolS over all six potentials. The average probability of these GCs' interactions is $\sim 10$\% on average in each of our six potentials.  

Of all the GCs, only a few, including NGC~2808, BH~140, Djorg~1, UKS~1, E~3, NGC~6656, and NGC~4833, have a statistically significant (10-20\%) probability of interaction with the SolS in at least one of the six potentials across all six billion years. The other GCs presented in the Table~\ref{tab:GC_SS} have a statistical probability from 5\% to 9\% over the six billion years. There were no close passages observed from the current day to 309~Myr in look-back time.

According to the possible progenitor classification of the GCs made by \cite{Massari2019} and \cite{Malhan2022}, 22~GCs can be from Gaia-Enceladus (G-E), seven can be from Pontus (ex situ), and only in a third place we have five GCs that are within the Galaxy.

Most of our selected GCs have tube (TB) orbits, 23, or long radial (LR) orbits, 17, and the other nine GCs have irregular (IR) type orbits. These orbital classifications were performed in \hyperlink{I23}{\color{blue}{Paper~I}}, according to Section~3.4 and Fig.~10 herein. 

Also, in Fig.~\ref{app:close_GC}, we present the Sun's orbital evolution with the collision 'points' with other GCs in the {\tt 411321} TNG external potential, which is presented in Table~\ref{tab:GC_SS}. The colours of the circles are more intense if the area has a higher statistical value for collisions. As can be seen in the top panel of the figure, the intense magenta colour represents a statistical probability of more than 10\% for NGC~7078. In the inner part of the Sun's orbit, according to the $R$, we observed that close passages are more frequent than in the outer part of the orbit.

\section{Influence of globular cluster gravity on the Oort cloud system} \label{sec:force}

For the Oort cloud, the initial equilibrium particle distribution is the well-established \citep[see Figure~9 in][]{Portegies2021b} Dehnen profile \citep{dehnen_family_1993} with power slope $\gamma = 2$: 
\begin{equation}\label{eq:rho_D}
    \rho_{\rm Oort}(r) = \frac{(3-\gamma) \cdot M_\mathrm{Oort}}{4\pi} \frac{a_\mathrm{Oort}}{r^\gamma (r+a_\mathrm{Oort})^{4-\gamma} },
\end{equation}
where M$_{\rm Oort}$ is the total mass of the model Oort cloud (in our case $0.01\rm\;M_{\odot}$), $a_\mathrm{Oort}$ is the Oort cloud scaling radius (which we set as $10^{5}$~au), and $\gamma$ describes the inner power-law profile of the Dehnen model family ($0\leq \gamma<3$). 


For the actual data generation, we used the AGAMA library \citep{agama2019}. The total mass of the Oort cloud we set initially equal to 1\% M$_\odot$ \citep{Portegies2021a}. The inner and outer cut-off radii were set at 100~au ($\sim 4.8 \times 10^{-4}$~pc) and 1~pc ($\sim 2 \times 10^{5}$~au). For the outer radius truncation, we simply used the Sun Hill radius definition in the Galactic potential (which is around $\sim$1 pc, as in the 
\citealt{Portegies2021a}). The particle distribution generated around the Sun only takes into account the Sun's gravitational field. As a basic model for our runs, we used $N = 10$k Oort cloud particles.

\begin{figure*}[htbp!]
\centering
\includegraphics[width=0.99\linewidth]{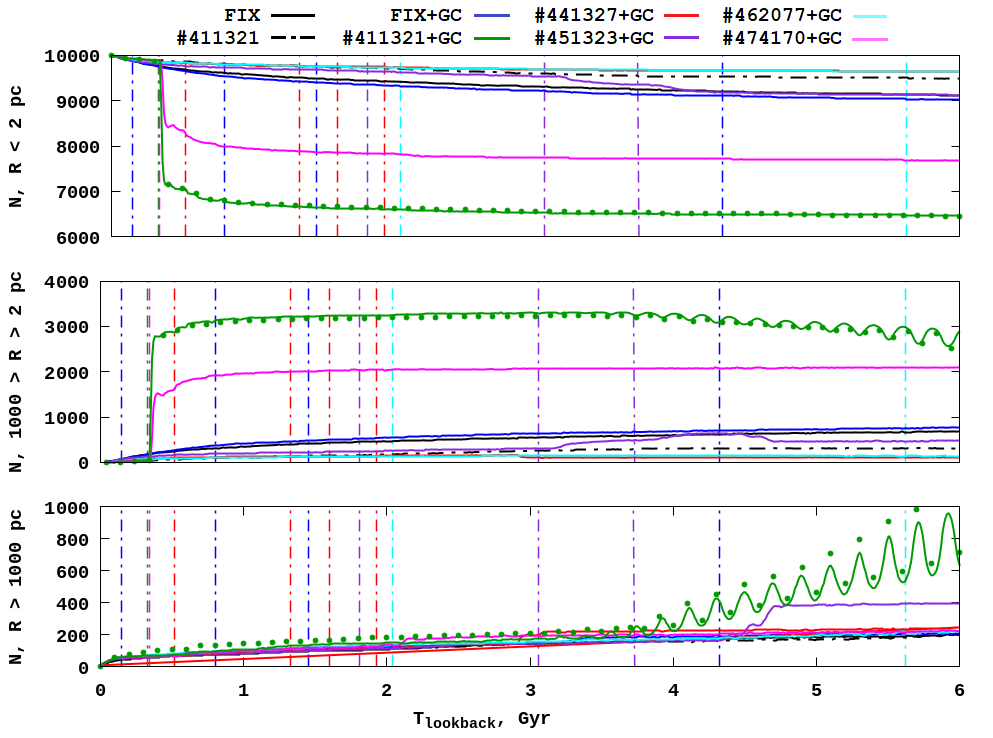}
\caption{Oort particle distributions due to gravitational influence of GCs. The different zones are as follows: Less than 2~pc -- inner zone (top panel), from 2~pc to 1000~pc -- middle zone (middle panel), and above 1000~pc -- outer zone (bottom panel). The solid lines show the particle distribution, and dotted lines show moments of the GC deep passages. The colours represent potentials. The larger green dots represent the results with the $N = 50$k Oort cloud number (re-scaled).}
\label{fig:pass_oort}
\end{figure*}

In the next step, we had cloud particles around the Sun and adopted the Oort cloud system in the five external TNG-TVP Milky Way-like and FIX potentials. For the whole system, we set the Sun's positions and velocities (see Section~\ref{sec:Intr}). Based on our Table~\ref{tab:GC_SS}, we selected the set of GCs with close passage distances from the Sun (less than 20~pc) in all 1000~sets of randomisation and for all potentials (five TNG-TVP and FIX). We modified our basic code to calculate the self-gravity only between the Sun and the GC as particles. We neglected the self-gravity of the Oort cloud particles in our investigation, following \cite{Portegies2021b}. As an independent check of Oort cloud dynamical evolution, we ran the cloud plus Sun system without the GC perturbers for the {\tt 411321} TNG-TVP and FIX potentials. On these runs, the Oort cloud loses only $\sim5$\% of particles (black lines in Fig.~\ref{fig:pass_oort}). 

In Table~\ref{tab:gc-pass}, we present the list of selected GCs that came closer than 10 pc to the Sun. In columns~(3) and (4) of the table, we present the minimal distance and time for when the GCs have a deep passage. To quantify the gravitational effect of GCs on the Oort cloud system, we defined $\varepsilon_{\rm TID,IN}$, which we set as a ratio of the GC potential to the Sun potential acting on a cloud system inner radius (column 7 in the table): 
\begin{equation}
\varepsilon_{\rm TID,IN} = \frac{\Phi_{\rm GC}}{\Phi_{\odot,\rm IN}}=
\frac{M_{\rm GC}\;[{\rm M_\odot}] /r_{\rm hm}\;[{\rm pc}]}{1\;[{\rm M_\odot}]/4.8 \cdot 10^{-4}\;{[\rm pc]}}, 
\label{eq:epsilon}
\end{equation}
where $M_{\rm GC}$ is the GC mass and $r_{\rm hm}$ is the GC half-mass radius (respectively columns 5 and 6 in Table~\ref{tab:gc-pass}).

In this way, we basically defined the gravitational energetic influence of the GCs acting on the whole Oort cloud system. To estimate the potential from the GC due to the deep passage of the SolS into the GC, we defined the GC potential based on the cluster $r_{\rm hm}$ instead of using a point mass approximation. 

We present a general overview of the interaction effect of the GCs in Fig.~\ref{fig:pass_oort}, which is related to Table~\ref{tab:gc-pass}. In this plot, for the analysis of the GC gravity influence on the Oort cloud particles, we established three zones based on the particle distances from the Sun: the inner part ($R$ is up to 2 pc), between ($R$ is from 2 to 1000 pc) and the outer part ($R$ is above 1000 pc). 

\begin{table}[htbp!]
\caption{Globular cluster physical parameters at the moment of deep close passages with the SolS.}
\centering
\begin{tabular}{llccccc}
\hline
\hline
Potential & GC & $dR$ & $T$ & $M_{\rm GC}$ & $r_{\rm hm}$ & $\varepsilon_{\rm TID,IN}$ \\
 & & pc & Gyr & $10^{5}\rm\;M_{\odot}$ & pc  & \\
 (1) & (2) & (3) & (4) & (5) & (6) & (7) \\
\hline
\hline
{\tt 411321} & NGC 7078 & 1.0      & 0.33 & 6.33 & 4.30 & 71$^\ast$ \\

{\tt 441327} & NGC 6426 &  2.1     & 0.52 & 0.71 & 8.00 & 4 \\
{\tt 441327} & UKS 1    &  2.3     & 1.33 & 0.77 & 3.84 & 10 \\
{\tt 441327} & Pal 10   &  4.5     & 1.60 & 1.62 & 6.33 & 12 \\
{\tt 441327} & NGC 6205 &  3.9     & 1.93 & 5.45 & 5.26 & 50 \\

{\tt 451323} & Pal 10   &  3.2     & 1.81 & 1.62 & 6.33 & 35 \\
{\tt 451323} & NGC 3201 &  0.9     & 3.06 & 1.60 & 6.78 & 11$^\ast$ \\
{\tt 451323} & NGC 2808 &  2.7     & 3.77 & 1.60 & 3.89 & 20$^\ast$ \\

{\tt 462077} & BH 140   &  4.1     & 2.04 & 0.59 & 9.53 & 3 \\
{\tt 462077} & NGC 7078 &  9.0     & 5.62 & 6.33 & 4.30 & 71 \\

{\tt 474170} & NGC 6356 &  1.2     & 0.34 & 6.00 & 6.86 & 42$^\ast$ \\

{\tt FIX} & UKS 1    &  1.3     & 0.15 & 0.77 & 3.84 & 10 \\
{\tt FIX} & BH 140   &  8.0     & 0.80 & 0.59 & 9.53 & 2 \\
{\tt FIX} & Djorg 1  &  5.5     & 1.45 & 0.79 & 5.57 & 7 \\
{\tt FIX} & Djorg 1  &  7.4     & 4.32 & 0.79 & 5.57 & 7 \\
\hline
\end{tabular}
\label{tab:gc-pass}
\tablefoot{
$^\ast$GCs that have a high gravitational influence on the Oort system; see Fig.~\ref{fig:pass_oort}.}
\end{table} 

As can be seen from Table~\ref{tab:gc-pass}, NGC~7078 has a high gravitational influence on the Oort system. Notably, the close passages of NGC~7078 are twice as deep as those of {\tt 411321} and {\tt 462077} TNG-TVPs. But in the first case, this GC comes as close as 1~pc to the Sun and as close as 9~pc in the second. As we observed from Fig.~\ref{fig:pass_oort}, at the moment NGC~7078 had a close passage in {\tt 411321} (marked by a green dotted line at 0.33~Gyr), more than $\sim30$\% of particles were immediately ejected from the first zone and moved to the second zone (green line). But in the case of {\tt 462077}, at the relative distance of 9~pc, NGC~7078 did not have any influence on the Oort system (cyan line, $T = 5.62$~Gyr).

Another example of the gravitational energetic influence of GCs on the Oort system is NGC~6356 in the {\tt 474170} external potential. At the 1.2~pc distance, we had $\varepsilon_{\rm TID,IN} = 42$ (magenta line, $T = 0.34$~Gyr), which caused $\sim20$\% of the particles to move from the first zone to the second. But in the third zone ($R$ above 1000~pc), this influence is almost negligible.   

In the {\tt 451323} external potential, we have three GCs with deep close passages, and only two of them have a soft influence on the Oort system, namely, NGC~3201 and NGC~2808 (violet lines, $T = 3.06$ and 3.77~Gyr). Their influence is perceptible only in the first and second zones.

The GCs UKS~1, BH~140, and NGC~6426, and others that have masses less than $\times10^{4}$ and have $dR$ less than 4 pc do not have any gravitational influence on the Oort system (see Fig.~\ref{fig:pass_oort}, blue and red dotted lines). The NGC~6205 with $\varepsilon_{\rm TID,IN} = 50$ in the {\tt 441327} potentially looks like an exception, though it does not have any effect on the Oort system either ($T = 1.93$~Gyr, red dotted line). 

We performed additional calculations to evaluate the possible influence of the initial particle number in the Oort system on the obtained results. In this case, we prepared the initial distribution of the Oort cloud particles with the same physical conditions as in a previous case, but we changed the number of particles, increasing it to 50k. Using these data, we carried out additional simulations with $N = 50$k for the FIX and {\tt 411321} potentials. The time evolution of the Oort particle numbers in different zones for $N = 10$k and 50k and for {\tt 411321} TNG-TVP we compare in Fig.~\ref{fig:pass_oort}, re-scaled (divided by five). The green line represents the $N = 10$k and the green dots represent the $N = 50$k. As can be seen, there are no significant differences between these two runs. We had an absolutely similar time when the NGC~7078 comes inside the Oort system, and as a result, we observed the ejection of more than 30\% of particles from the first zone to the second zone. After four billion years in look-back time, we observed small differences in the number of particles in the second and third zones that clearly reflect the result of the small particle numbers (especially in the third zone). We present the detailed distribution of the Oort cloud particles' gravitational potentials for six billion years in look-back time in Fig.~\ref{fig:fit_oort}. 

\section{Conclusions}\label{sec:con}

After analysing the total interaction of GCs with the SolS, we found that close passages are not frequent but are also not exceptional events in the SolS lifetime. One can note that besides a quite different nature in the time variable of TNG-100 and FIX external potentials, the resulting close passage interaction of GCs with the SolS for these external potentials are surprisingly similar. For example, at a relative distance of 50~pc between a GC and the SolS, we obtained on average $\sim 15$\% of the close passage probability over all six billion years, and at $dR=100$~pc, we got on average $\sim 35$\% of the close passage probability over all six billion years. 

Summarising our set of calculations in Table~\ref{tab:gc-pass} and Fig.~\ref{fig:pass_oort}, we observed that a potentially strong gravitational influence from deep close passages as close as a 1--2 pc on the Oort system would yield objects with typical masses above a few times $10^{5} ~{\rm M}_{\odot}$. Estimating the potential influence of the GCs on the SolS Oort clouds, we should expect that the GCs in the past could have had much higher masses and half-mass radii that would obviously lead to the stronger effects on the SolS.

\begin{acknowledgements}
The authors thank the anonymous referee for a very constructive report and suggestions that helped significantly improve the quality of the manuscript.

MI and PB acknowledge the support within the grant No.~AP14869395 of the Science Committee of the Ministry of Science and Higher Education of Kazakhstan (``Triune model of Galactic center dynamical evolution on cosmological time scale'').

MI, PB, and MS thanks the support from the special program of the Polish Academy of Sciences and the U.S. National Academy of Sciences under the Long-term program to support Ukrainian research teams grant No.~ PAN.BFB.S.BWZ.329.022.2023.

The work of MI was supported by the Grant of the National Academy of Sciences of Ukraine for young scientists. 

MS acknowledges the support under the Fellowship of the President of Ukraine for young scientists 2022-2024.

\end{acknowledgements}

\bibliographystyle{mnras}  
\bibliography{part_4}   

\begin{thebibliography}{}
\makeatletter
\relax
\def\mn@urlcharsother{\let\do\@makeother \do\$\do\&\do\#\do\^\do\_\do\%\do\~}
\def\mn@doi{\begingroup\mn@urlcharsother \@ifnextchar [ {\mn@doi@}
  {\mn@doi@[]}}
\def\mn@doi@[#1]#2{\def\@tempa{#1}\ifx\@tempa\@empty \href
  {http://dx.doi.org/#2} {doi:#2}\else \href {http://dx.doi.org/#2} {#1}\fi
  \endgroup}
\def\mn@eprint#1#2{\mn@eprint@#1:#2::\@nil}
\def\mn@eprint@arXiv#1{\href {http://arxiv.org/abs/#1} {{\tt arXiv:#1}}}
\def\mn@eprint@dblp#1{\href {http://dblp.uni-trier.de/rec/bibtex/#1.xml}
  {dblp:#1}}
\def\mn@eprint@#1:#2:#3:#4\@nil{\def\@tempa {#1}\def\@tempb {#2}\def\@tempc
  {#3}\ifx \@tempc \@empty \let \@tempc \@tempb \let \@tempb \@tempa \fi \ifx
  \@tempb \@empty \def\@tempb {arXiv}\fi \@ifundefined
  {mn@eprint@\@tempb}{\@tempb:\@tempc}{\expandafter \expandafter \csname
  mn@eprint@\@tempb\endcsname \expandafter{\@tempc}}}

\bibitem[\protect\citeauthoryear{{Bailer-Jones}}{{Bailer-Jones}}{2022}]{Bailer-Jones2022}
{Bailer-Jones} C.~A.~L.,  2022, \mn@doi [\apjl] {10.3847/2041-8213/ac816a},
  \href {https://ui.adsabs.harvard.edu/abs/2022ApJ...935L...9B} {935, L9}

\bibitem[\protect\citeauthoryear{{Baumgardt} \& {Vasiliev}}{{Baumgardt} \&
  {Vasiliev}}{2021}]{Baumgardt2021}
{Baumgardt} H.,  {Vasiliev} E.,  2021, \mn@doi [\mnras]
  {10.1093/mnras/stab1474}, \href
  {https://ui.adsabs.harvard.edu/abs/2021MNRAS.505.5957B} {505, 5957}

\bibitem[\protect\citeauthoryear{{Bennett} \& {Bovy}}{{Bennett} \&
  {Bovy}}{2019}]{Bennett2019}
{Bennett} M.,  {Bovy} J.,  2019, \mn@doi [\mnras] {10.1093/mnras/sty2813},
  \href {https://ui.adsabs.harvard.edu/abs/2019MNRAS.482.1417B} {482, 1417}

\bibitem[\protect\citeauthoryear{{Bennett}, {Bovy}  \& {Hunt}}{{Bennett}
  et~al.}{2022}]{Bennett2022}
{Bennett} M.,  {Bovy} J.,   {Hunt} J. A.~S.,  2022, \mn@doi [\apj]
  {10.3847/1538-4357/ac5021}, \href
  {https://ui.adsabs.harvard.edu/abs/2022ApJ...927..131B} {927, 131}

\bibitem[\protect\citeauthoryear{{Bland-Hawthorn} \&
  {Gerhard}}{{Bland-Hawthorn} \& {Gerhard}}{2016}]{Bland-Hawthorn2016}
{Bland-Hawthorn} J.,  {Gerhard} O.,  2016, \mn@doi [\araa]
  {10.1146/annurev-astro-081915-023441}, \href
  {https://ui.adsabs.harvard.edu/abs/2016ARA&A..54..529B} {54, 529}

\bibitem[\protect\citeauthoryear{{Bonanno}, {Schlattl}  \&
  {Patern{\`o}}}{{Bonanno} et~al.}{2002}]{Bonanno2002}
{Bonanno} A.,  {Schlattl} H.,   {Patern{\`o}} L.,  2002, \mn@doi [\aap]
  {10.1051/0004-6361:20020749}, \href
  {https://ui.adsabs.harvard.edu/abs/2002A&A...390.1115B} {390, 1115}

\bibitem[\protect\citeauthoryear{{Connelly}, {Bizzarro}, {Krot}, {Nordlund},
  {Wielandt}  \& {Ivanova}}{{Connelly} et~al.}{2012}]{Connelly2012S}
{Connelly} J.~N.,  {Bizzarro} M.,  {Krot} A.~N.,  {Nordlund} {\r{A}}.,
  {Wielandt} D.,   {Ivanova} M.~A.,  2012, \mn@doi [Science]
  {10.1126/science.1226919}, \href
  {https://ui.adsabs.harvard.edu/abs/2012Sci...338..651C} {338, 651}

\bibitem[\protect\citeauthoryear{{Dehnen}}{{Dehnen}}{1993}]{dehnen_family_1993}
{Dehnen} W.,  1993, \mn@doi [\mnras] {10.1093/mnras/265.1.250}, \href
  {https://ui.adsabs.harvard.edu/abs/1993MNRAS.265..250D} {265, 250}

\bibitem[\protect\citeauthoryear{{Gravity Collaboration} et~al.,}{{Gravity
  Collaboration} et~al.}{2019}]{Gravity2019}
{Gravity Collaboration} et~al., 2019, \mn@doi [\aap]
  {10.1051/0004-6361/201935656}, \href
  {https://ui.adsabs.harvard.edu/abs/2019A&A...625L..10G} {625, L10}

\bibitem[\protect\citeauthoryear{{Ishchenko}, {Sobolenko}, {Berczik},
  {Khoperskov}, {Omarov}, {Sobodar}  \& {Makukov}}{{Ishchenko}
  et~al.}{2023a}]{Ishchenko2023a}
{Ishchenko} M.,  {Sobolenko} M.,  {Berczik} P.,  {Khoperskov} S.,  {Omarov} C.,
   {Sobodar} O.,   {Makukov} M.,  2023a, \mn@doi [\aap]
  {10.1051/0004-6361/202245117}, \href
  {https://ui.adsabs.harvard.edu/abs/2023A&A...673A.152I} {673, A152
  \hypertarget{I23}{(Paper~I)}}

\bibitem[\protect\citeauthoryear{{Ishchenko}, {Sobolenko}, {Kuvatova},
  {Panamarev}  \& {Berczik}}{{Ishchenko} et~al.}{2023b}]{Ishchenko2023b}
{Ishchenko} M.,  {Sobolenko} M.,  {Kuvatova} D.,  {Panamarev} T.,   {Berczik}
  P.,  2023b, \mn@doi [\aap] {10.1051/0004-6361/202245753}, \href
  {https://ui.adsabs.harvard.edu/abs/2023A&A...674A..70I} {674, A70}

\bibitem[\protect\citeauthoryear{{Ishchenko}, {Sobolenko}, {Berczik}, {Omarov},
  {Sobodar}, {Kalambay}  \& {Yurin}}{{Ishchenko}
  et~al.}{2023c}]{Ishchenko2023c}
{Ishchenko} M.,  {Sobolenko} M.,  {Berczik} P.,  {Omarov} C.,  {Sobodar} O.,
  {Kalambay} M.,   {Yurin} D.,  2023c, \mn@doi [\aap]
  {10.1051/0004-6361/202346553}, \href
  {https://ui.adsabs.harvard.edu/abs/2023A&A...678A..69I} {678, A69}

\bibitem[\protect\citeauthoryear{{Johnson} \& {Soderblom}}{{Johnson} \&
  {Soderblom}}{1987}]{Johnson1987}
{Johnson} D. R.~H.,  {Soderblom} D.~R.,  1987, \mn@doi [\aj] {10.1086/114370},
  \href {https://ui.adsabs.harvard.edu/abs/1987AJ.....93..864J} {93, 864}

\bibitem[\protect\citeauthoryear{{Just}, {Berczik}, {Petrov}  \&
  {Ernst}}{{Just} et~al.}{2009}]{Just2009}
{Just} A.,  {Berczik} P.,  {Petrov} M.~I.,   {Ernst} A.,  2009, \mn@doi
  [\mnras] {10.1111/j.1365-2966.2008.14099.x}, \href
  {https://ui.adsabs.harvard.edu/abs/2009MNRAS.392..969J} {392, 969}

\bibitem[\protect\citeauthoryear{{Karim} \& {Mamajek}}{{Karim} \&
  {Mamajek}}{2017}]{Karim2017}
{Karim} T.,  {Mamajek} E.~E.,  2017, \mn@doi [\mnras] {10.1093/mnras/stw2772},
  \href {https://ui.adsabs.harvard.edu/abs/2017MNRAS.465..472K} {465, 472}

\bibitem[\protect\citeauthoryear{{Khoperskov}, {Mastrobuono-Battisti}, {Di
  Matteo}  \& {Haywood}}{{Khoperskov} et~al.}{2018}]{Khoperskov2018}
{Khoperskov} S.,  {Mastrobuono-Battisti} A.,  {Di Matteo} P.,   {Haywood} M.,
  2018, \mn@doi [\aap] {10.1051/0004-6361/201833534}, \href
  {https://ui.adsabs.harvard.edu/abs/2018A&A...620A.154K} {620, A154}

\bibitem[\protect\citeauthoryear{{Malhan} et~al.,}{{Malhan}
  et~al.}{2022}]{Malhan2022}
{Malhan} K.,  et~al., 2022, \mn@doi [\apj] {10.3847/1538-4357/ac4d2a}, \href
  {https://ui.adsabs.harvard.edu/abs/2022ApJ...926..107M} {926, 107}

\bibitem[\protect\citeauthoryear{{Mardini} et~al.,}{{Mardini}
  et~al.}{2020}]{Mardini2020}
{Mardini} M.~K.,  et~al., 2020, \mn@doi [\apj] {10.3847/1538-4357/abbc13},
  \href {https://ui.adsabs.harvard.edu/abs/2020ApJ...903...88M} {903, 88}

\bibitem[\protect\citeauthoryear{{Massari}, {Koppelman}  \& {Helmi}}{{Massari}
  et~al.}{2019}]{Massari2019}
{Massari} D.,  {Koppelman} H.~H.,   {Helmi} A.,  2019, \mn@doi [\aap]
  {10.1051/0004-6361/201936135}, \href
  {https://ui.adsabs.harvard.edu/abs/2019A&A...630L...4M} {630, L4}

\bibitem[\protect\citeauthoryear{{Miyamoto} \& {Nagai}}{{Miyamoto} \&
  {Nagai}}{1975}]{Miyamoto1975}
{Miyamoto} M.,  {Nagai} R.,  1975, \pasj, \href
  {https://ui.adsabs.harvard.edu/abs/1975PASJ...27..533M} {27, 533}

\bibitem[\protect\citeauthoryear{{Nelson} et~al.,}{{Nelson}
  et~al.}{2018}]{Nelson2018}
{Nelson} D.,  et~al., 2018, \mn@doi [\mnras] {10.1093/mnras/stx3040}, \href
  {https://ui.adsabs.harvard.edu/abs/2018MNRAS.475..624N} {475, 624}

\bibitem[\protect\citeauthoryear{{Nelson} et~al.,}{{Nelson}
  et~al.}{2019a}]{Nelson2019}
{Nelson} D.,  et~al., 2019a, \mn@doi [Computational Astrophysics and Cosmology]
  {10.1186/s40668-019-0028-x}, \href
  {https://ui.adsabs.harvard.edu/abs/2019ComAC...6....2N} {6, 2}

\bibitem[\protect\citeauthoryear{{Nelson} et~al.,}{{Nelson}
  et~al.}{2019b}]{NelsonPill2019}
{Nelson} D.,  et~al., 2019b, \mn@doi [\mnras] {10.1093/mnras/stz2306}, \href
  {https://ui.adsabs.harvard.edu/abs/2019MNRAS.490.3234N} {490, 3234}

\bibitem[\protect\citeauthoryear{{Pfalzner} \& {Govind}}{{Pfalzner} \&
  {Govind}}{2021}]{Pfalzner2021a}
{Pfalzner} S.,  {Govind} A.,  2021, \mn@doi [\apj] {10.3847/1538-4357/ac19aa},
  \href {https://ui.adsabs.harvard.edu/abs/2021ApJ...921...90P} {921, 90}

\bibitem[\protect\citeauthoryear{{Pfalzner} \& {Vincke}}{{Pfalzner} \&
  {Vincke}}{2020}]{Pfalzner2020}
{Pfalzner} S.,  {Vincke} K.,  2020, \mn@doi [\apj] {10.3847/1538-4357/ab9533},
  \href {https://ui.adsabs.harvard.edu/abs/2020ApJ...897...60P} {897, 60}

\bibitem[\protect\citeauthoryear{{Pfalzner}, {Aizpuru Vargas}, {Bhandare}  \&
  {Veras}}{{Pfalzner} et~al.}{2021}]{Pfalzner2021}
{Pfalzner} S.,  {Aizpuru Vargas} L.~L.,  {Bhandare} A.,   {Veras} D.,  2021,
  \mn@doi [\aap] {10.1051/0004-6361/202140587}, \href
  {https://ui.adsabs.harvard.edu/abs/2021A&A...651A..38P} {651, A38}

\bibitem[\protect\citeauthoryear{{Portegies Zwart}}{{Portegies
  Zwart}}{2021}]{Portegies2021a}
{Portegies Zwart} S.,  2021, \mn@doi [\aap] {10.1051/0004-6361/202038888},
  \href {https://ui.adsabs.harvard.edu/abs/2021A&A...647A.136P} {647, A136}

\bibitem[\protect\citeauthoryear{{Portegies Zwart}, {Torres}, {Cai}  \&
  {Brown}}{{Portegies Zwart} et~al.}{2021}]{Portegies2021b}
{Portegies Zwart} S.,  {Torres} S.,  {Cai} M.~X.,   {Brown} A. G.~A.,  2021,
  \mn@doi [\aap] {10.1051/0004-6361/202040096}, \href
  {https://ui.adsabs.harvard.edu/abs/2021A&A...652A.144P} {652, A144}

\bibitem[\protect\citeauthoryear{{Reid} \& {Brunthaler}}{{Reid} \&
  {Brunthaler}}{2004}]{Reid2004}
{Reid} M.~J.,  {Brunthaler} A.,  2004, \mn@doi [\apj] {10.1086/424960}, \href
  {https://ui.adsabs.harvard.edu/abs/2004ApJ...616..872R} {616, 872}

\bibitem[\protect\citeauthoryear{{Sch{\"o}nrich}, {Binney}  \&
  {Dehnen}}{{Sch{\"o}nrich} et~al.}{2010}]{Schonrich2010}
{Sch{\"o}nrich} R.,  {Binney} J.,   {Dehnen} W.,  2010, \mn@doi [\mnras]
  {10.1111/j.1365-2966.2010.16253.x}, \href
  {https://ui.adsabs.harvard.edu/abs/2010MNRAS.403.1829S} {403, 1829}

\bibitem[\protect\citeauthoryear{{Shukirgaliyev} et~al.,}{{Shukirgaliyev}
  et~al.}{2021}]{Shukirgaliyev2021}
{Shukirgaliyev} B.,  et~al., 2021, \mn@doi [\aap]
  {10.1051/0004-6361/202141299}, \href
  {https://ui.adsabs.harvard.edu/abs/2021A&A...654A..53S} {654, A53}

\bibitem[\protect\citeauthoryear{{Vasiliev}}{{Vasiliev}}{2019}]{agama2019}
{Vasiliev} E.,  2019, \mn@doi [\mnras] {10.1093/mnras/sty2672}, \href
  {https://ui.adsabs.harvard.edu/abs/2019MNRAS.482.1525V} {482, 1525}

\bibitem[\protect\citeauthoryear{{de la Fuente Marcos} \& {de la Fuente
  Marcos}}{{de la Fuente Marcos} \& {de la Fuente Marcos}}{2022}]{Marcos2022}
{de la Fuente Marcos} R.,  {de la Fuente Marcos} C.,  2022, \mn@doi [Research
  Notes of the American Astronomical Society] {10.3847/2515-5172/ac842b}, \href
  {https://ui.adsabs.harvard.edu/abs/2022RNAAS...6..152D} {6, 152}

\bibitem[\protect\citeauthoryear{{de la Fuente Marcos}, {de la Fuente Marcos}
  \& {Reilly}}{{de la Fuente Marcos} et~al.}{2014}]{Marcos2014}
{de la Fuente Marcos} R.,  {de la Fuente Marcos} C.,   {Reilly} D.,  2014,
  \mn@doi [\apss] {10.1007/s10509-013-1635-7}, \href
  {https://ui.adsabs.harvard.edu/abs/2014Ap&SS.349..379D} {349, 379}

\makeatother
\end{thebibliography}

\begin{appendix}
\section{List of GCs with close passages near the Solar System} \label{app:table_GC}

In Table~\ref{tab:GC_SS}, we present the mean minimum relative distance ($<dR_{\rm m}>$) (minimum relative distance between GCs and SolS from each of our 6000~randomisations) and relative velocity ($<dV>$) (also averaged by all randomisations). Column~(5) of the table shows the first close passage end the relative distance at this moment, $<T_{\rm l}> \Leftrightarrow <dR>$. In Columns~(6) and (7), we present the GC's current mass and half-mass radii. Column~(8) shows the gravity influence on the Oort cloud system (rounded to the closest integer). To quantify the gravitational effect of a GC on the Oort cloud system, we re-defined the $\varepsilon_{\rm TID,IN}$ (see equation~\ref{eq:epsilon}), which we set as a ratio of the GC potential to the Sun potential acting on a cloud system inner radius: 
\begin{equation}
\epsilon_{\rm TID,IN} = \frac{\Phi_{\rm GC}}{\Phi_{\odot,\rm IN}}=
\frac{M_{\rm GC}\;[{\rm M_\odot}] /dR_{\rm m}\;[{\rm pc}]}{1\;[{\rm M_\odot}]/4.8 \cdot 10^{-4}\;{[\rm pc]}}, 
\label{eq:epsilon2}
\end{equation}
where now we use the $dR_{\rm m}$ values from column~(3) of the table and assume that the GC acts as a particle on the inner part of the Oort cloud system. Columns~(9) and (10) show the type of GC orbits (TO) and different regions of the Galaxy (GR) at present, according to the classification from \cite{Bland-Hawthorn2016}. Column~(11) shows the classification of the progenitors according to the \cite{Massari2019} and \cite{Malhan2022}.

\begin{table*}[!b]
\caption{Globular clusters that have close passages near the Solar System}
\centering
\resizebox{0.9\textwidth}{!}{
\begin{tabular}{lcccccccrcc}
\hline
\hline
\multicolumn{1}{c}{ID} & GC & {$<dR_{\rm m}>$} & {$<dV>$} &  {$<T_{\rm l}> \Leftrightarrow <dR>$} & \multicolumn{1}{c}{$M_{\rm GC}$} & \multicolumn{1}{c}{$R_{\rm hm}$} & $\epsilon_{\rm TID,IN}$ & TO & GR & Progenitor \\
 &  & pc & km~s$^{-1}$ & Gyr $\Leftrightarrow$ pc & \multicolumn{1}{c}{{$10^{5}\rm\;M_{\odot}$}} & \multicolumn{1}{c}{pc} & & & \\
  (1) & (2) & (3) & (4) & (5) & (6) & (7) & (8) & (9) & (10) & (11) \\
\hline
\hline
1  & NGC~362    & 70    & 309 & 2.48 $\Leftrightarrow$ 151 & 2.84 &  3.79 & 2 & LR & HL  & Pontus \\  
2  & NGC 288    & 53.5  & 343 & 2.30 $\Leftrightarrow$ 129 & 0.93 &  8.37 & 1 & IR & HL  & Pontus \\ 
3  & NGC~1261   & 60.4  & 402 & 3.31 $\Leftrightarrow$ 93 & 1.82 &  5.23 & 1 & IR & HL  & G-E    \\ 
4  & Pal~2      & 31    & 434 & 1.26 $\Leftrightarrow$ 114 & 2.31 &  8.06 & 4 & LR & HL  & G-E   \\ 
5  & \textit{NGC~1851}   & 48.5 & 401 & 1.62 $\Leftrightarrow$ 128 & 3.18 & 2.90 & 3 & LR & HL & G-E  \\ 
6  & \textit{NGC~1904}   & 38.3 & 373 & 1.56 $\Leftrightarrow$ 113 & 1.39 & 3.21 & 2 & LR & HL & G-E \\  
7  & NGC~2298   & 65.3  & 364 & 1.71 $\Leftrightarrow$ 129 & 5.58 & 0.33 & 4 & IR & HL  & G-E    \\  
8  & \textit{NGC~2808} & 38.7  & 288 & 1.85 $\Leftrightarrow$ 148 & 8.64 &  3.89 & 8 & TB & HL & G-E  \\ 
9  & ESO~280-06 & 29.1  & 297 & 0.73 $\Leftrightarrow$ 135 & 0.78 &  0.97 & 1 & TB & HL  & --     \\  
10 & Rup~106    & 122   & 340 & 3.37 $\Leftrightarrow$ 150 & 3.42 & 1.16 & 1 & TB & HL  & H99   \\  
11 &  \textit{\textbf{BH~140}}     & 9.65  & 196 &  1.021 $\Leftrightarrow$ 132 &  0.6 &  9.53 & 3 & TB & HL & -- \\
12 & NGC~5286   & 50.8  & 335 & 1.04 $\Leftrightarrow$ 129 & 0.35 &  3.79 & 0 & TB & HL  & Pontus \\ 
13 & NGC~5634   & 84.6  & 379 & 5.02 $\Leftrightarrow$ 100 & 2.28 &  7.39 & 1 & IR & HL  & G-E    \\
14 & NGC~5694   & 92.6  & 569 & 2.43 $\Leftrightarrow$ 132 & 3.17 &  4.86 & 2 & LR & HL  & H-E   \\
15 & Pal~5      & 79.4  & 311 & 2.57 $\Leftrightarrow$ 144 & 0.1 & 27.64 & 0 & IR & HL  & LMS 1  \\
16 & NGC~5897   & 20.8  & 241 & 1.51 $\Leftrightarrow$ 138 & 1.57 & 10.99 & 4 & TB & HL  & G-E   \\
17 & NGC~6205   & 65.8  & 299 & 2.60 $\Leftrightarrow$ 98 & 5.45 &  5.26 & 4 & TB & HL  & Pontus \\
18 & NGC~6229   & 78.4  & 417 & 1.64 $\Leftrightarrow$ 118 & 2.86 &  4.41 & 2 & LR & HL  & G-E    \\ 
19 & Pal~15     & 127   & 463 & 2.10 $\Leftrightarrow$ 136 & 0.51 & 26.86 & 0 & LR & HL & G-E$^\ast$  \\
20 & NGC~6341   & 43    & 280 & 2.03 $\Leftrightarrow$ 145 & 3.52 &  4.49 & 4 & LR & HL & Pontus \\ 
21 & NGC~6356  & 32.8  & 188 & 0.92 $\Leftrightarrow$ 157 & 6.00 &  6.86 & 9 & TB & HL & G-D \\ 
22 & IC~1257    & 65.1  & 397 &  1.028 $\Leftrightarrow$ 123 &  0.18 &  5.59 & 0 & LR & HL & G-E \\
23 & FSR~1758  & 36.5  & 373 & 0.79 $\Leftrightarrow$ 127 & 6.28 & 17.04 & 8 & TB & HL & Seq \\
24 &  \textit{\textbf{Djorg~1}}   & 19.67 & 231 & 0.309 $\Leftrightarrow$ 144 & 0.78 &  5.57 & 2 & TB & HL & G-E \\  
25 & NGC~6584   & 101  & 327 & 2.66 $\Leftrightarrow$ 135 & 1.02 &  5.37 & 0 & IR & HL & G-E \\
26 &  \textit{\textbf{UKS~1}}     & 17.55 & 239 & 0.40 $\Leftrightarrow$ 151 & 0.77 &  3.84 & 2 & TB & HL & -- \\ 
27 & NGC~6981   & 78.9  & 385 & 2.21 $\Leftrightarrow$ 163 &  0.69 &  5.96 & 0 & LR & HL & G-E \\  
28 & NGC~6779   & 45.3  & 353 & 1.32 $\Leftrightarrow$ 118 & 1.86 &  4.51 & 2 & TB & HL & Pontus \\ 
29 & Pal~10     & 23.7  & 101 & 0.69 $\Leftrightarrow$ 134 & 1.62 &  6.33 & 3 & TB & HL & G-D \\
30 & Pal~11     & 33.6  & 149 & 0.84 $\Leftrightarrow$ 149 & 0.12 &  7.72 & 0 & TB & HL & G-D \\
31 & NGC~6864   & 45.1  & 337 & 1.16 $\Leftrightarrow$ 148 & 3.70 &  2.96 & 4 & TB & HL & G-E \\
32 & NGC~7006   & 96.2  & 536 & 2.25 $\Leftrightarrow$ 141 & 1.36 &  6.99 & 1 & LR & HL & Seq \\
33 & \textit{NGC~7078}  & 24.5 & 180 & 2.73 $\Leftrightarrow$ 139 & 6.33  &  4.30 & 12 & TB & HL & G-D \\
34 & NGC~7099  & 38.6 & 300 & 2.63 $\Leftrightarrow$ 146 & 1.43 &  4.99 & 2 & TB & HL & Pontus \\
35 & NGC~7492  & 70.2 & 434 & 2.93 $\Leftrightarrow$ 108 & 0.26 &  9.89 & 0 & LR & HL & G-E \\  
\hline
36 & \textit{E~3}        & 59.4  & 139 & 1.135 $\Leftrightarrow$ 133 & 0.03 & 27.14 & 0 & TB & HL & H99  \\ 
37 & \textit{NGC~3201}   & 28.5  & 566 & 3.70 $\Leftrightarrow$ 172 & 1.60  & 6.78  & 3 & TB & HL & Arjuna/Sequoia \\ 
38 & NGC~4147   & 127   & 406 & 2.92 $\Leftrightarrow$ 134 & 3.90  & 0.40  & 1 & LR & HL & G-E \\ 
39 & NGC~6235   & 106   & 193 & 1.91 $\Leftrightarrow$ 149 & 10.7  & 0.48  & 5 & TB & HL & G-E \\ 
40 & \textit{NGC~6426}   & 54.1  & 254 & 2.67 $\Leftrightarrow$ 140 & 0.72  & 8.00  & 1 & TB & HL & H-E  \\ 
41 & \textit{NGC~6656}   & 39.3  & 187 & 2.46 $\Leftrightarrow$ 107 & 4.76  & 5.29  & 6 & TB & HL & M-D  \\ 
42 & NGC~6934   & 133   & 381 & 3.42 $\Leftrightarrow$ 110 & 1.36 & 5.16  & 0 & LR & HL & H-E  \\ 
43 & NGC~7089   & 70    & 380 & 2.80 $\Leftrightarrow$ 120 & 6.33 & 4.30  & 4 & LR & HL & G-E  \\ 
\hline
44 & Eridanus   & 129   & 511 & 1.902 $\Leftrightarrow$ 148 & 0.12  & 17.91 & 0 & IR & HL & H-E  \\ 
45 & IC~4499    & 80.6  & 491 & 4.256 $\Leftrightarrow$ 91 & 1.55  & 14.96 & 1 & IR & HL & Seq  \\ 
46 & NGC~5024   & 97.2  & 374 & 4.32 $\Leftrightarrow$ 141 & 4.55  & 17.31 & 2 & IR & HL & LMS-1/Wukong \\ 
47 & \textit{NGC~4833}   & 60.7 & 223 & 2.28 $\Leftrightarrow$ 114 & 2.06  & 4.76  & 2 & TB & HL & G-E  \\ 
48 & NGC~5904   & 69.7  & 377 & 355 $\Leftrightarrow$ 180 & 3.92 & 0.57  & 3 & LR & HL & H99/G-E \\ 
49 & Pal~4      & 145   & 550 & 1.45 $\Leftrightarrow$ 157 & 0.13  & 21.30 & 0 & LR & HL & G-E \\ 
\hline
\end{tabular}
}
\tablefoot{The bold text indicates the GCs that have a minimum distance of close passages. 
The italic text indicates the GCs that have a statistical probability of close passages near the Sun of more than 10\%.}
\label{tab:GC_SS}
\end{table*}

\section{Orbital evolution of the Sun with the collision moments} \label{app:close_GC}
As an example, we present the orbital evolution of the Sun with collision moments for several GCs: BH~140, UKS~1, Djorg~1, NGC~7078, NGC~2808. The colour points represents the counterparts of GC's during the whole six-billion-years time interval.

\begin{figure*}[htbp!]
\centering
\includegraphics[width=0.99\linewidth]{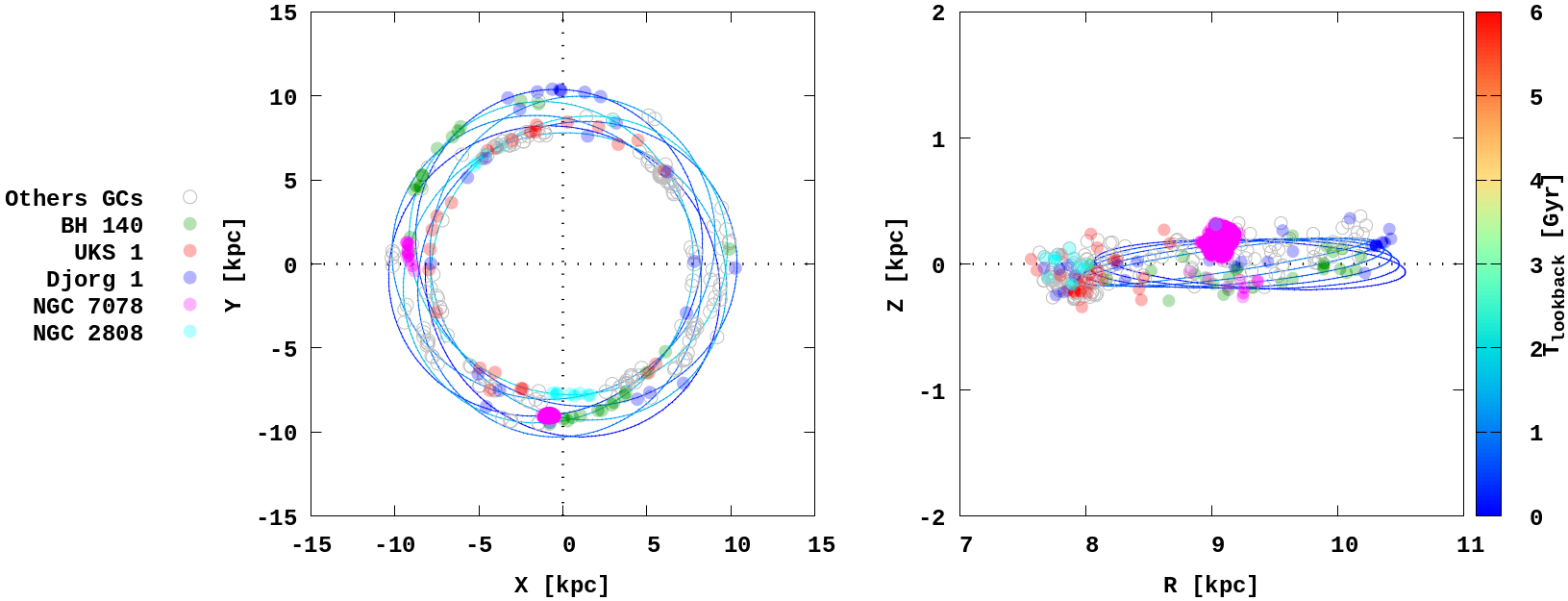}
\includegraphics[width=0.99\linewidth]{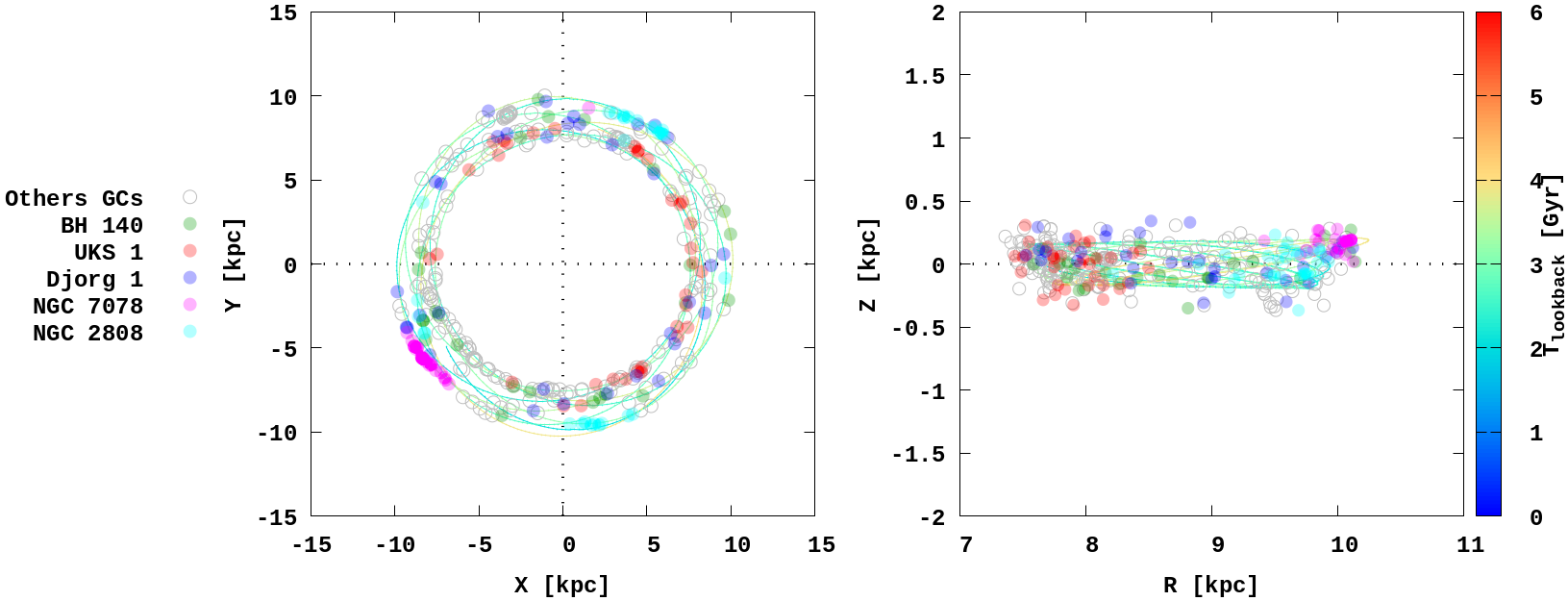}
\includegraphics[width=0.99\linewidth]{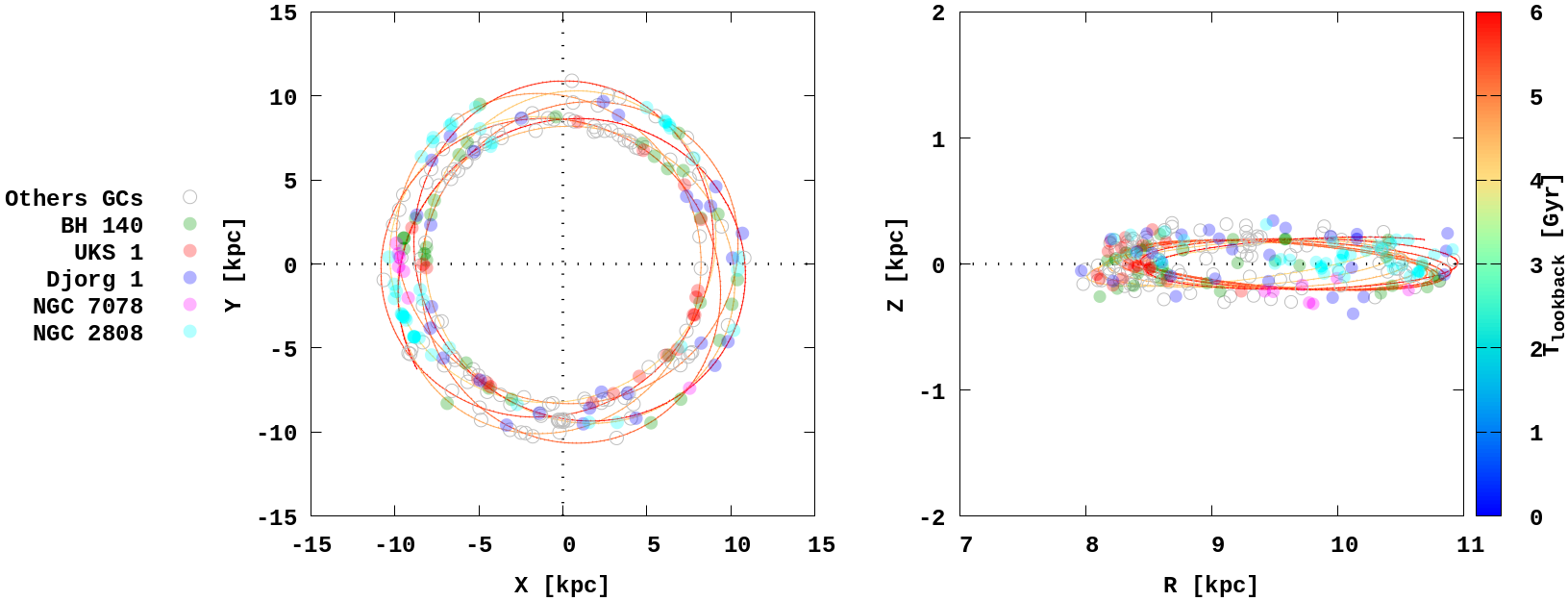}
\caption{Orbital Sun evolution with close passages during the whole integration time in {\tt 411321} TNG-TVP. Coloured circles: green for BH~140, red for UKS~1, blue for Djorg~1, magenta for NGC~7078, cyan for NGC~2808. The hollow circles outlined in grey are other GCs. The blue point is the start of integration. From top to bottom, the panels represent the interval of integration: 0 -- 2~Gyr, 2 -- 4~Gyr, and 4 -- 6~Gyr in look-back time.}
\label{fig:sun_orb}
\end{figure*}

\section{Oort particle distributions due to the GCs gravitational influence} \label{app:POT}

\begin{figure*}[htbp!]
\centering
\includegraphics[height=0.30\linewidth]{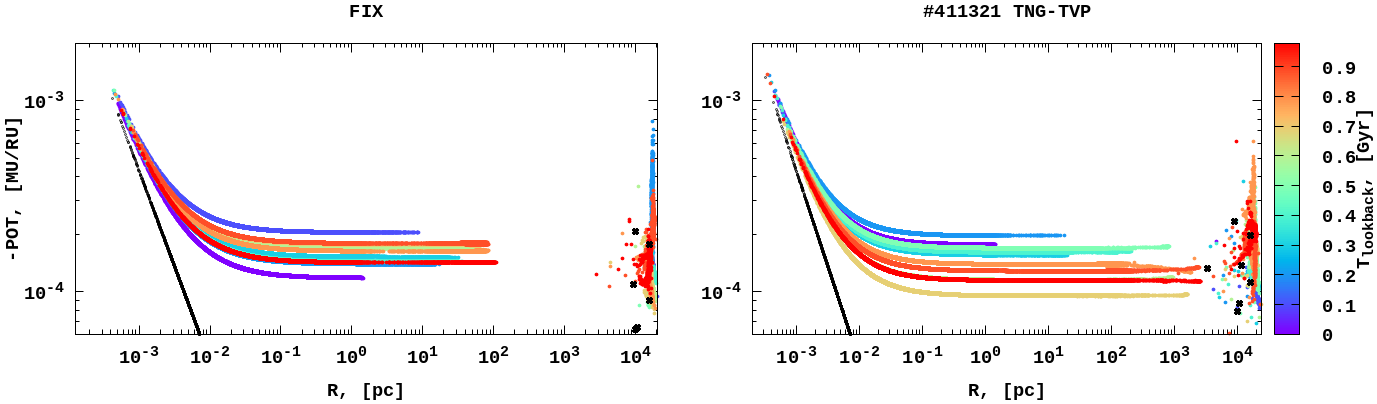}
\includegraphics[height=0.30\linewidth]{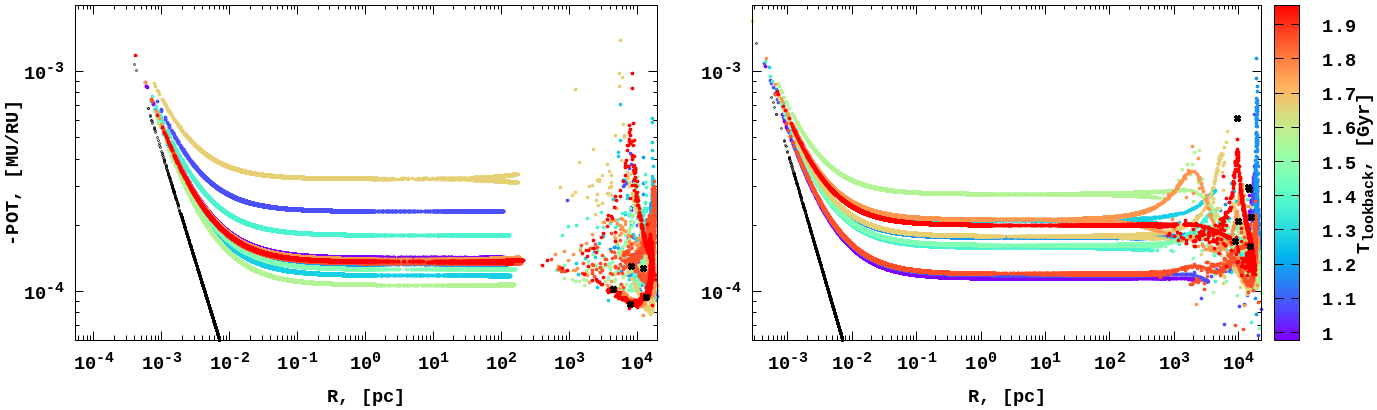}
\includegraphics[height=0.30\linewidth]{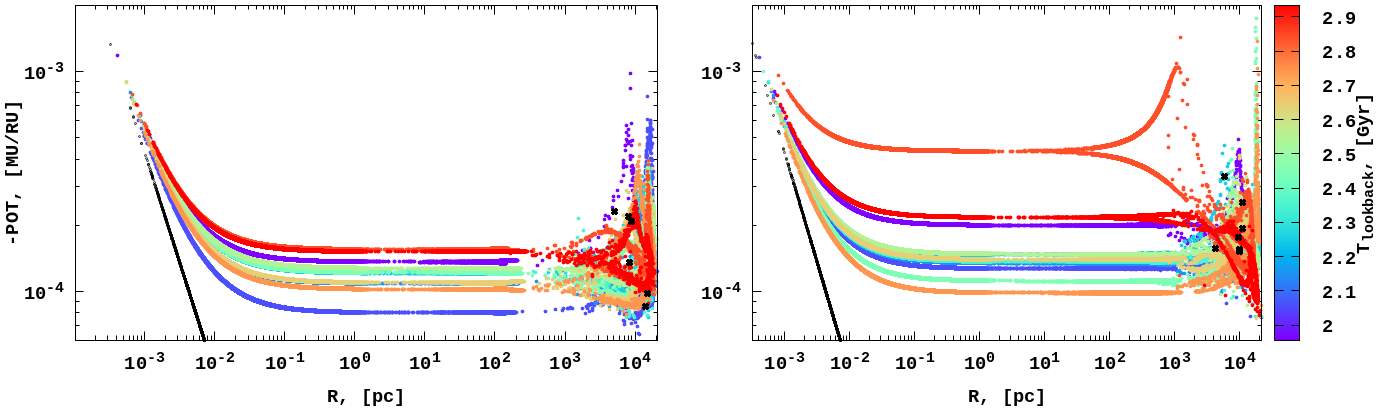}
\caption{Values of the gravitational potential (with '-' sign) for each Oort particle in FIX and {\tt 411321} potentials for the first three billion years. Black lines are particle potentials without the GC influence, and colour lines are the same but with the GC influence. Each panel represents one billion years in look-back time integration. The FIX potential is presented in the left panels, and {\tt 411321} is on the right panels.}
\label{fig:fit_oort}
\end{figure*}

In Fig.~\ref{fig:fit_oort} we present the gravitational potential acting on each Oort particle for each one-billion year interval for each panel with 100~million-year time steps (coloured lines). The black dots on the right edge of the panels show the position of the GCs at the beginning for each set of a billion years. The black lines represent the values of the gravitational potential for each Oort particle due to only the Sun for the FIX and {\tt 411321} potentials. This gravitational potential from the Sun can be described as a simple function $\sim R^{-1}$. For the potential normalisation, we used the N-body units of global integration: G = 1, MU = 2.325 $\cdot$ 10$^9$ M$_\odot$, RU = 1 kpc, VU = 100 km/s. 

The coloured lines represent the distribution of the Oort particles in the FIX and {\tt 411321} potentials due to the gravitational influence of the GCs. In these cases, the influence of the GCs on the Oort system is more complex and has a pronounced flat part in outer radii, $R > 2$~pc. In the right panels, we observed that the influence of the GCs in the case of {\tt 411321} TNG-TVP is stronger, and many more stripped former Oort particles become bound with the flyby GCs. In a late phase of the evaluation (after $\sim3$~Gyr), we observed a strong separation between the different families of the Oort particle orbits. The gap formation at the $R= 2$~pc is directly connected with the process of the Oort cloud particles' strong tidal stripping due to the direct gravitational influence of the GCs. The captured particles at $R > 1$~kpc are under the strong gravitational influence of the GCs. In this case, their gravitational potential is no longer connected with the Sun but is mainly defined by the influence of the closest GC.

\begin{figure*}[htbp!]
\centering
\includegraphics[height=0.30\linewidth]{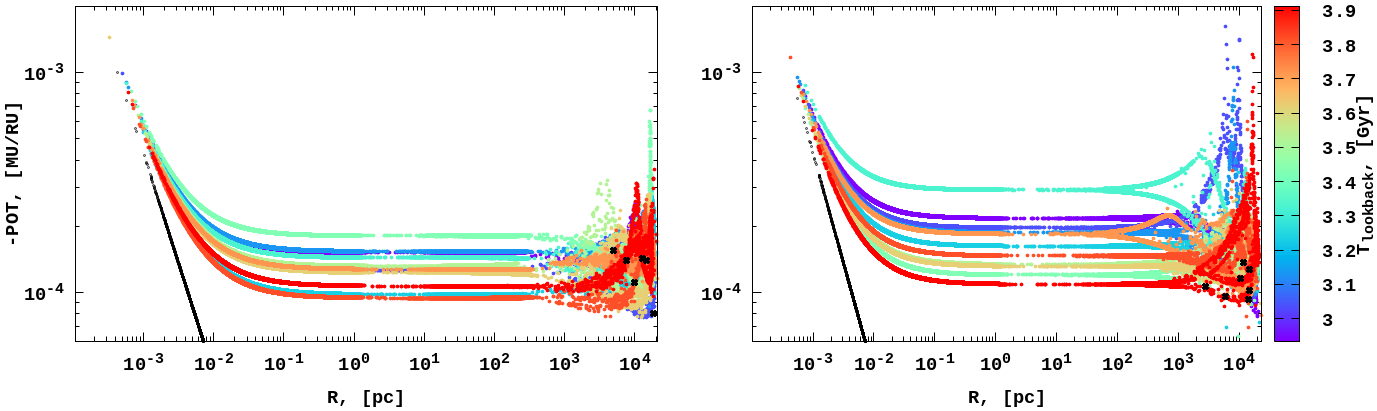}
\includegraphics[height=0.30\linewidth]{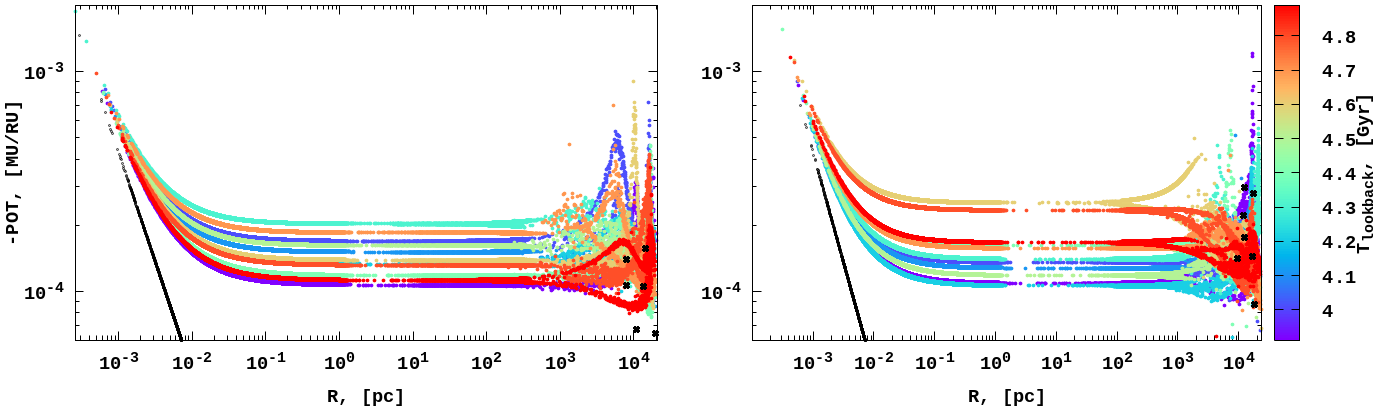}
\includegraphics[height=0.30\linewidth]{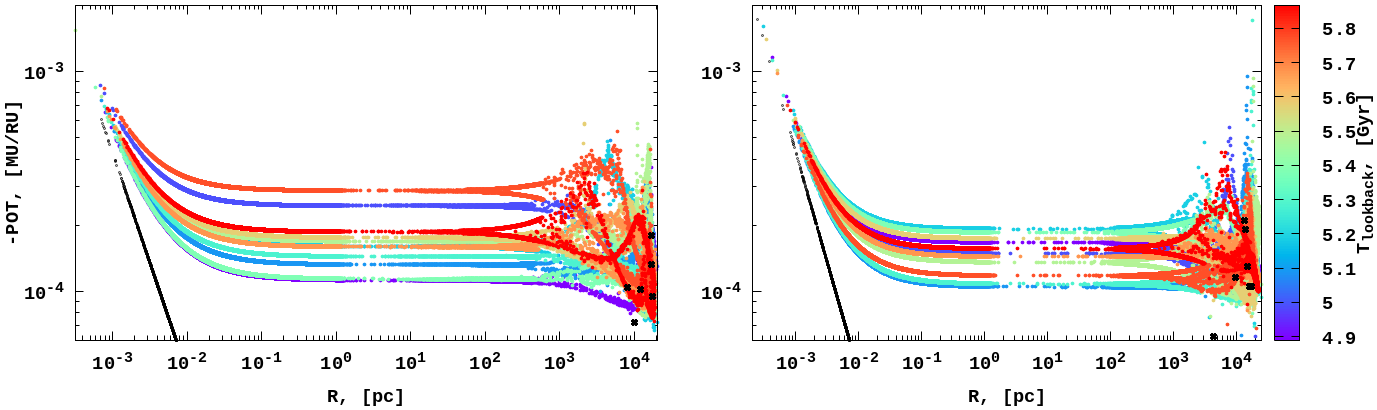}
\caption{Same as in Fig.~\ref{fig:fit_oort} but for the 3--6 Gyr interval.}
\label{fig:fit_oort1}
\end{figure*}

\end{appendix}

\end{document}